\documentclass[journal]{vgtc}                
\ifpdf
  \pdfoutput=1\relax                   
  \usepackage{graphicx}                
  \DeclareGraphicsExtensions{.pdf,.png,.jpg,.jpeg} 
\else
  \usepackage{graphicx}                
  \DeclareGraphicsExtensions{.eps}     
\fi%

\graphicspath{{figures/}{pictures/}{images/}{./}} 

\usepackage{microtype}                 
\PassOptionsToPackage{warn}{textcomp}  
\usepackage{textcomp}                  
\usepackage{mathptmx}                  
\usepackage{times}                     
\usepackage{cite}                      
\usepackage{tabu}                      
\usepackage{booktabs}                  
\usepackage{paralist}
\usepackage{multirow}
\usepackage{enumitem}

\usepackage[dvipsnames]{xcolor}
\definecolor{hcolor}{HTML}{fffae6}

\usepackage[titlenumbered, ruled, noend]{algorithm2e}
\usepackage{scalerel}

\usepackage{tikz}
\newcommand*\circled[1]{\tikz[baseline=(char.base)]{
            \node[shape=circle,draw,inner sep=0.5pt] (char) {#1};}}

\newcommand{\name}{SightBi}

\newcommand{\eg}{e.g.}
\newcommand{\ie}{i.e.}

\def\markup{1}
\if\markup1
\newcommand{\rv}[1]{{\leavevmode\color{Black}#1}}
\else
\newcommand{\rv}[1]{#1}
\fi

\onlineid{1293}

\vgtccategory{Research}
\vgtcpapertype{Representations \& Interaction}

\title{\name: Exploring Cross-View Data Relationships with Biclusters}

\author{Maoyuan Sun, Abdul Rahman Shaikh, Hamed Alhoori, Jian Zhao}
\authorfooter{
\item
Maoyuan Sun (corresponding author), Abdul Rahman Shaikh, and Hamed Alhoori are with Northern Illinois University. E-mail: \{smaoyuan, ashaikh2, alhoori\}@niu.edu.
\item
Jian Zhao is with University of Waterloo. E-mail: jianzhao@uwaterloo.ca.
}

\shortauthortitle{Biv \MakeLowercase{\textit{et al.}}: Global Illumination for Fun and Profit}
\abstract{%
Multiple-view visualization (MV) has been heavily used in visual analysis tools for sensemaking of data in various domains (e.g., bioinformatics, cybersecurity and text analytics).
One common task of visual analysis with multiple views is to relate data across different views.
For example, to identify threats, an intelligence analyst needs to link people from a social network graph with locations on a crime-map, and then search for and read relevant documents.
Currently, exploring cross-view data relationships heavily relies on view-coordination techniques (e.g., brushing and linking), which may require significant user effort on many trial-and-error attempts, such as repetitiously selecting elements in one view, and then observing and following elements highlighted in other views.
\rv{To address this, we present \name, a visual analytics approach for supporting cross-view data relationship explorations.
We discuss the design rationale of \name\ in detail, with identified user tasks regarding the use of cross-view data relationships.
\name\ formalizes cross-view data relationships as biclusters, computes them from a dataset, and uses a bi-context design that highlights creating stand-alone relationship-views.
This helps preserve existing views and offers an overview of cross-view data relationships to guide user exploration.
Moreover, \name\ allows users to interactively manage the layout of multiple views by using newly created relationship-views.
With a usage scenario, we demonstrate the usefulness of \name\ for sensemaking of cross-view data relationships.}
} 


\CCScatlist{ 
 \CCScat{K.6.1}{Management of Computing and Information Systems}%
{Project and People Management}{Life Cycle};
 \CCScat{K.7.m}{The Computing Profession}{Miscellaneous}{Ethics}
}

\teaser{
   \centering
   \includegraphics[width=\linewidth]{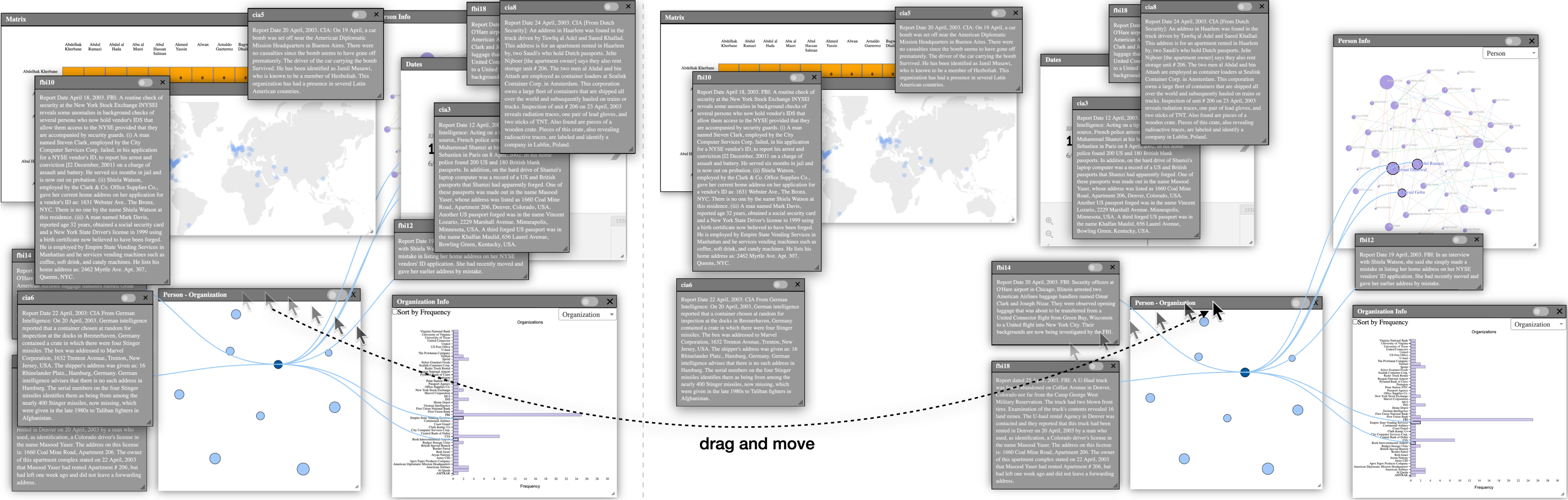}
   \caption{\name\ shows computed cross-view data relationships in stand-alone relationship-views and supports users in flexibly organizing the layout of multiple views. 
   As a user drags a relationship-view, its connected views automatically move with it. 
   This enables a user to move a set of views out of a cluttered layout and helps preserve a meaningful spatialization previously created by the user.}
   \label{fig-teaser}
 }

\vgtcinsertpkg

\begin{document}



\firstsection{Introduction}

\maketitle

\textit{Multiple-view visualization} (MV) has been heavily used for sensemaking of data.
Similar to a multi-focus approach \cite{zhao2015interactive}, MV highlights that each view supports certain analysis tasks by showing data in a specific type of visualization. 
Different views either show different parts of data or display the same data from different perspectives as different visualizations.
MV has been incorporated in visual analysis tools in various fields, such as Caleydo \cite{lex2010caleydo} for biomolecular data analysis, Canopy \cite{burtner2013interactive} for multimedia analysis, IN-SPIRE \cite{inspire} and Jigsaw \cite{stasko2008jigsaw} for text analytics, and Tableau \cite{tableau} and Spotfire \cite{ahlberg1996spotfire} for business intelligence.

\begin{figure}[th]
  \centering
  \includegraphics[width=\columnwidth]{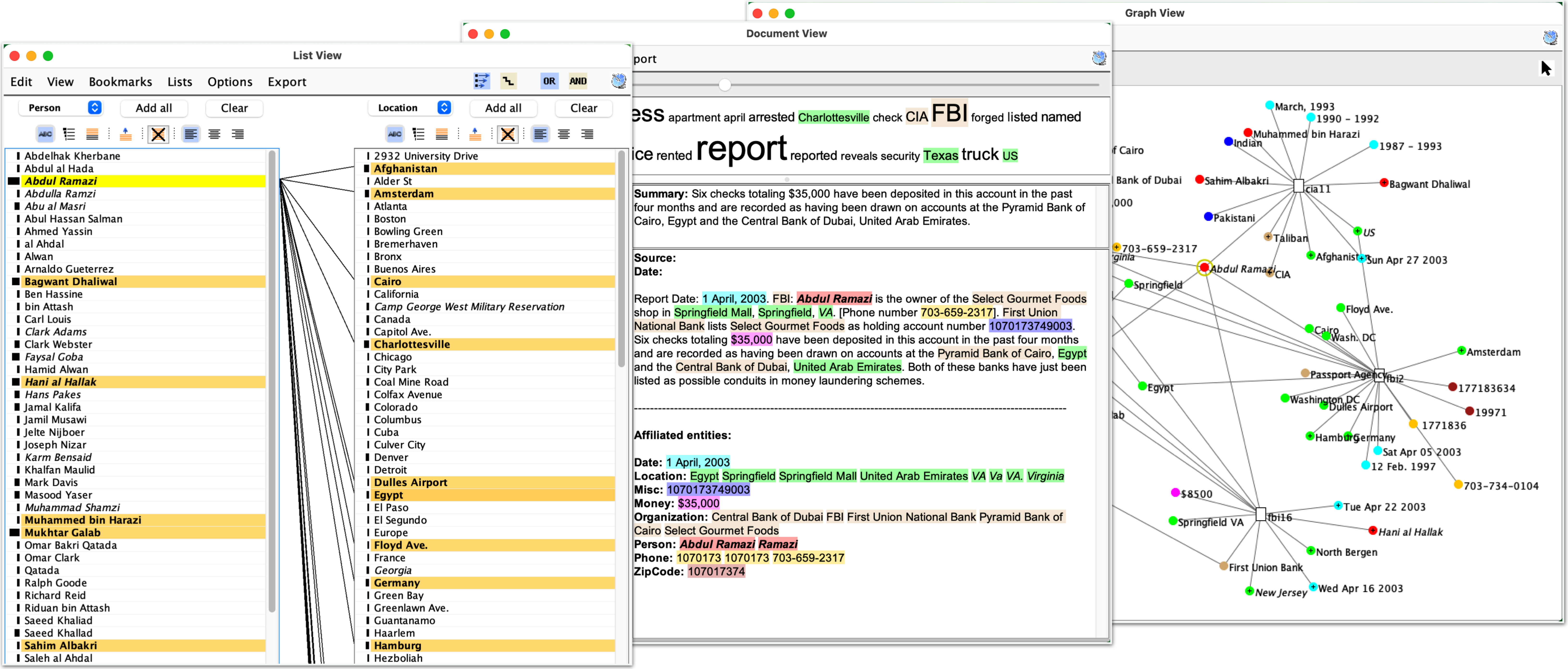}
    \vspace{-6mm}
  \caption{An example of relating data from three views in Jigsaw: a list view (left), a document view (middle), and a graph view (right).
  }
  ~\label{fig-jigsaw}
    \vspace{-5mm}
\end{figure}

To gain a comprehensive understanding of data, analysts often need to relate data from multiple views.
For example, as is shown in Figure \ref{fig-jigsaw}, in a text analytics scenario, an analyst, Sarah, explores intelligence reports to identify threats by using Jigsaw \cite{stasko2008jigsaw}.
After loading the dataset, Sarah works on three views offered by Jigsaw: a list view showing connections between entities, a document view displaying text, and a graph view presenting links between entities and documents.
Sarah tries to relate their information by exploring entity relations, following connections between entities and documents, and reading relevant text.
With Jigsaw's view coordination functions, when Sarah clicks on an entity in the list view, corresponding entities and documents are highlighted in other views.
Sarah needs to repetitiously click entities in one view and follow highlighted ones in other views to explore various possible combinations of related data across the three views.
As the number of clicked entities grows, Sarah soon gets confused about which selected entities in the list view connect to which pieces of highlighted text in the document view (e.g., are they all related or are there only partial connections between some of them).


Exploring \textit{cross-view data relationships} is not as simple as it looks like.
Currently, MV-based visual analysis tools heavily rely on view-coordination techniques \cite{north1997taxonomy, roberts2007state, pattison2001view, wang2000guidelines, boukhelifa2003coordination} for users to relate data from different views.
Since these techniques offer limited visual guidance, users have to put significant amount of effort into many trial-and-error attempts (e.g., trying to select different entities in a view, following and checking highlighted parts in other views).
This potentially forces users to manually solve a combinatorial problem: discovering sets of entities in each view and requiring that entities in such sets from different views are related \rv{(e.g., finding sets of nodes in a social network graph that are related to sets of locations on a map and sets of organizations in a list)}.

Computation can help save human effort in finding connected sets of entities between different domains (\eg, person and location).
Specifically, \textit{biclustering} \cite{madeira2004biclustering} has been developed to simultaneously cluster two related sets of entities into related subsets.
Such a related pair of subsets is a \textit{bicluster}, where entities in one set are related to those in another.
Based on shared entities, biclusters can be linked together to form \textit{bicluster-chains}.
Figure \ref{fig-bic} shows examples of biclusters and bicluster-chains.
With them, we can compute data relationships between a pair of views or among several views.
While using biclusters helps free users from a time-consuming trial-and-error process, it brings challenging problems. 
How can we apply biclustering to MV, especially when the number of views goes beyond two? 
How can we visualize biclusters in MV to support users in exploring cross-view data relationships, but not affecting much of existing views?


To address such challenges, we propose \name, a novel visual analytics technique for exploring cross-view data relationships.
With a relational data model, \name\ computes data relationships between pairs of views as biclusters, and shows them in newly created relationship-views.
Moreover, \name\ allows users to steer cross-view data relationships computing by interactively including or excluding views, which drives the process of chaining multiple biclusters together.
In summary, this work highlights the following three contributions.


1) We formalize cross-view data relationships as biclusters and present a data model to support this. 
It enables computing data relationships between pairs of views and supports flexibly including or excluding views for cross-view data relationship exploration.

2) We propose a \textit{bi-context} design concept to show computed cross-view data relationships. 
It highlights separating cross-view data relationships from existing views by creating stand-alone views for computed relationships. 
The newly added relationship-views help preserve existing views and can serve as overviews to guide user exploration.

3) We develop a visualization prototype, \name. 
It implements the proposed concept and supports users in interactively exploring data relationships across multiple views.
We demonstrate the usefulness of \name\ 
with an investigative analytics scenario.

\section{Related Work}

\subsection{Relating Information from Multiple Views}
\label{mview-prior-work}

Three major designs support relating data across multiple views: 1) \textit{linkage}, 2) \textit{coordination}, and 3) \textit{proximity}.

\textbf{Linkage} is a straightforward approach that maps data relations across multiple views to \textit{visible links}. 
It offers the most explicit way of showing cross-view data relationships. 
By following a visible link, users can see a cross-view relationship and investigate its involved components. 
This design concept has been applied to show connections between visual elements in different views, such as Bixplorer \cite{sun2014role}, ConnectedCharts \cite{viau2012connectedcharts}, GPLOM \cite{im2013gplom}, Flowstrates \cite{boyandin2011flowstrates}, ImAxes \cite{cordeil2017imaxes}, MyBrush \cite{koytek2018mybrush}, PivotSlice \cite{zhao2013interactive}, Semantic Substrates \cite{shneiderman2006network}, VisLink \cite{collins2007vislink}, and variant versions of VisLink \cite{waldner2010visual, geymayer2014show, waldner2011collaborative}.
As each visible link associates two components, it can only reveal a one-to-one relationship. 
When the number of relationships is large, this design may cause visual clutter (e.g., many links crossing each other).
For complex relationships (e.g., many-to-many relationships), such visible links cannot directly reveal them, so users have to manually trace linked elements, identify shared ones, and then group some links together.

\textbf{Coordination} is a relatively implicit strategy to show cross-view data relationships. 
It highlights a \textit{dynamic updating} approach supported by necessary user interactions (e.g., brushing).
Cross-view relationships will not be revealed until users interact with some visual elements.
User interactions trigger some changes in visual encodings in multiple views. 
Users need to refer to such changes to understand relationships.
Coordination has been heavily explored and used in visual analysis tools (e.g., Cross-filtered Views \cite{weaver2010cross}, CViews \cite{boukhelifa2003model}, Improvise \cite{weaver2004building}, Snap-Together \cite{north2000snap}, and Spotfire \cite{ahlberg1996spotfire}).
However, coordination suffers from two problems. 
First, 
it requires users to pay enough attention to changes in multiple views corresponding to their interactions. 
Users may not realize cross-view relationships if they fail to recognize changes after their interactions. 
Second, 
detailed connections can hardly be revealed by coordination. 
For example, when users brush five nodes in a scatterplot, seven bars get highlighted in a histogram. 
Users can learn that these five nodes and the seven bars are related, but whether or not each node is related to all the bars cannot be clearly answered.

\textbf{Proximity} 
focuses on an \textit{arrangement oriented} approach. 
It highlights spatially organizing relationship components. 
Proximity is less explicit than the other two. 
Users cannot directly see related visual elements across multiple views. 
Instead, they have to understand organizations of relationship components to explore cross-view relationships. 
Proximity can be implemented in a spatial organization that uses relative distances in a 2D space to indicate potential relationships between views. 
For example, users place documents A and B near each other to indicate that they have similar topics.
It has been used to support sensemaking \cite{pirolli2005sensemaking} (e.g., Analyst’s Workstation \cite{andrews2010space}, Bixplorer \cite{fiaux2013bixplorer}, Co-Cluster Analysis \cite{xu2016interactive}, ForceSPIRE \cite{endert2012semantic} and NodeTrix \cite{henry2007nodetrix}).
Proximity neither suffers from visual clutter issues caused by visible links, nor relies on users perceiving changes to recognize relationships.
Yet, to understand cross-view relationships, proximity requires users to create, investigate or reason about spatialization, which needs much cognitive effort.
\rv{Moreover, it works for small sets of views in a 2D space. 
For a large number of views, proximity may not work due to overlap.}

\begin{figure}[tb]
  \centering
  \includegraphics[width=\columnwidth]{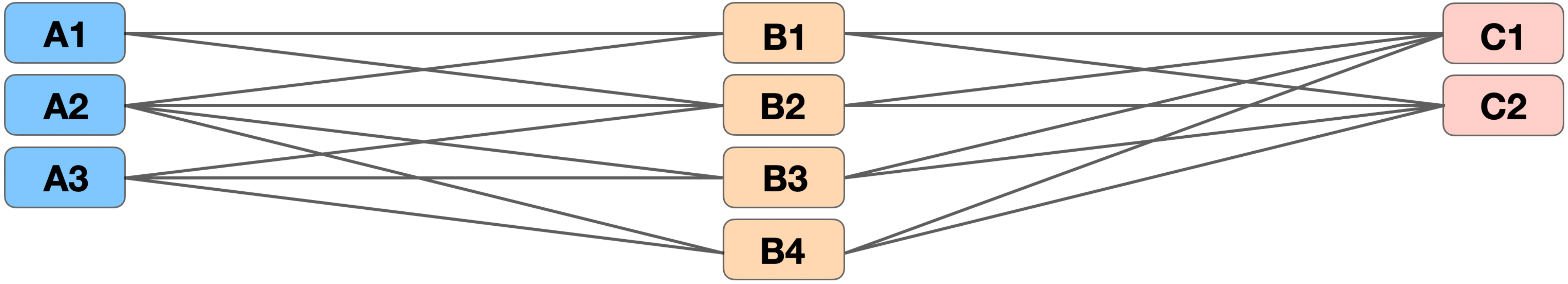}
    \vspace{-6mm}
  \caption{Three biclusters: 1) $\{A1, A2\}-\{B1, B2\}$, 2) $\{A2, A3\}-\{B2, B3, B4\}$, and 3) $\{B1, B2, B3, B4\}-\{C1, C2\}$. 1) and 2) share $A2$ and $B2$. 1) and 3) share $B1$ and $B2$. 2) and 3) share $B2, B3$ and $B4$. With shared entities, there are two bicluster-chains, composed of 1) and 3), and 2) and 3).
  }
  ~\label{fig-bic}
    \vspace{-5mm}
\end{figure}

\subsection{Bicluster and Bicluster-Chain}
\label{bic}



Biclusters are results from biclustering algorithms \cite{madeira2004biclustering}, which simultaneously compute subsets of entities and subsets of conditions where the set of entities behave similarly under the set of conditions.
It has been applied to solve real-world problems, such as discovering co-behaved genes in bioinformatics\cite{madeira2004biclustering}, identifying bundled shopping items in marketing analysis \cite{uno2004efficient}, finding colluding threats in intelligence analysis \cite{wu2018interactive}, and detecting useful missing edges for network analysis \cite{zhao2020understanding, zhao2019missbin}. 
From a graph perspective, biclusters are bicliques (\ie, complete bipartite graphs), where each vertex in one set is linked to all vertices in another set (see Figure \ref{fig-bic}).
In data mining, algorithms (\eg, LCM \cite{uno2004efficient} and CHARM \cite{zaki2005efficient}) are typically designed for computing \textit{closed biclusters}, which are maximal bicliques.

Biclusters can overlap by sharing entities \cite{sun2016biset}.
Based on shared entities, biclusters, consisting of entities from multiple domains, can be linked to form \textit{bicluster-chains} \cite{sun2014five}.
Figure \ref{fig-bic} shows two bicluster-chains: \rv{$\{A1, A2\}-\{B1, B2\}-\{C1, C2\}$ and $\{A2, A3\}-\{B2, B3, B4\}-\{C1, C2\}$. 
The former is based on shared entities, $B1$ and $B2$. 
The latter is based on shared entities, $B2, B3$ and $B4$.}
Different biclusters can share different entities, so it is possible to use the number of shared entities as the level of overlap for finding bicluster-chains \cite{zhao2017bidots}. 

\subsection{Bicluster Visualizations}
To make computed biclusters usable, bicluster visualizations have been studied. 
Sun et al. \cite{sun2014five} proposed a design framework for bicluster visualizations. 
It considers designs for five relationship-levels: \textit{entity}-level, \textit{group}-level, \textit{bicluster}-level, \textit{chain}-level, and \textit{schema}-level. 
Entity-level refers to a one-to-one relationship, and group-level is a one-to-many relationship.
Bicluster-level and chain-level are many-to-many relationships that involve two or more sets of entities from different domains (\eg, person, location and date).
Schema-level regards the relational schema of a database.

A key design trade-off of visualizing biclusters has been identified: \textit{preserving the entity uniqueness} vs. \textit{maintaining the cluster completeness} \cite{sun2016visual}. 
The former requires that visualizing biclusters without duplicating entities. 
The latter highlights that placing entities of the same biclusters neighboring each other.
Due to overlap, it is impossible to achieve both for certain visual layouts (e.g., lists and matrices).
Fundamentally, this is an Euler diagram problem for visualizing multiple sets \cite{alsallakh2016state}.
Based on the trade-off, there are three designs for visualizing biclusters: \textit{entity}-centric, \textit{relationship}-centric, and \textit{cluster}-centric.

An entity-centric design focuses on preserving the uniqueness of entities, with a hypothesis that entity duplication may confuse users \cite{sun2016biset}.
It has been applied to a node-link diagram based layout (e.g., multiple lists or node-link graphs).
In a multi-list based layout, biclusters are perceived by tracing edges (e.g., finding bicliques in Jigsaw's List view \cite{stasko2008jigsaw}). 
In a node-link graph, biclusters can be revealed with set visualization techniques (e.g., Bubble Sets \cite{collins2009bubble}, LineSets \cite{alper2011design} and Kelpfusion \cite{meulemans2013kelpfusion}) by creating an additional layer on top of a graph.
Users need to trace edges or follow added set-layers to identify a bicluster, because entities spread out over node-link diagrams.

A relationship-centric design emphasizes the relationship between entities.
It aims to place entities of the same biclusters near each other. 
It often picks a matrix as the basic layout to visualize biclusters (\eg, Bicluster viewer \cite{heinrich2011bicluster}, BiVoc \cite{grothaus2006automatic}, and BicOverlapper \cite{santamaria2014bicoverlapper}), where rows are entities in one set and columns are another set (\eg, a \textsl{person}-\textsl{location} relationship matrix). 
To show biclusters, it requires users to order rows and columns, so that entities of the same biclusters can be near each other. 
This generates sub-matrices inside a relationship matrix to show biclusters.
Due to overlaps, one order can not show all biclusters.
To see all biclusters, users need to change the matrix order, or some rows or columns in the matrix have to be duplicated \cite{streit2014furby}.

A  cluster-centric design encodes each bicluster with its own visual mark. 
To achieve this, entities shared by multiple biclusters are duplicated.
It has been used in hybrid visualizations that replace nodes in a node-link diagram with some charts.
For example, in Bixplorer \cite{fiaux2013bixplorer} and Furby \cite{streit2014furby}, each bicluster is shown as a matrix. 
If two biclusters share entities, two matrices are linked with edges.
BiDot \cite{zhao2017bidots} uses bipartite graphs as its basic layout and replaces nodes in the graph with two sets of aligned dots to show biclusters.
For each bicluster, the identified trade-off is balanced, as entities are gathered without duplication. 
However, when the number is above one, the more biclusters overlap, the more entities are duplicated.
To address this, BiSet \cite{sun2016biset} encodes each bicluster as an edge bundle in a multi-list layout and supports users in interactively ordering entities \cite{sun2018effect} and merging biclusters \cite{sun2019interactive} in lists.
It uses marks for biclusters out of entity-lists, which preserves existing entity-lists. 
This separation inspired the design of the relationship-view in \name, because we aim to maintain existing views.

\section{Definition and Specification}

\begin{figure}[tb]
  \centering
  \includegraphics[width=\columnwidth]{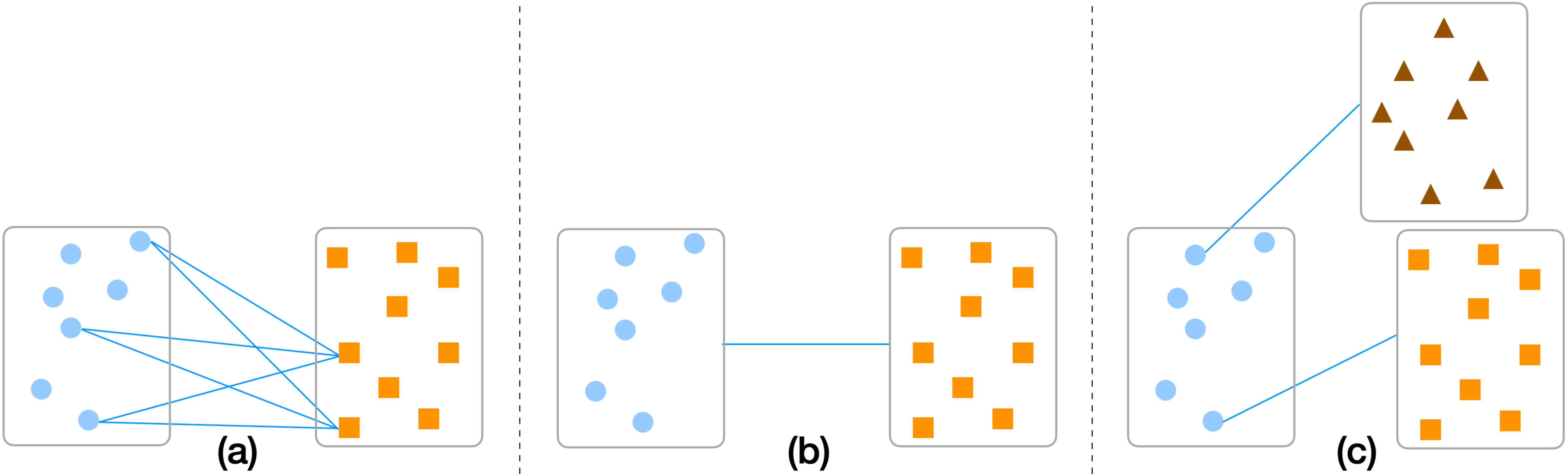}
    \vspace{-6mm}
  \caption{Three types of cross-view relationships: (a) between visual elements, (b) between views and (c) between visual elements and views. 
  }
  ~\label{3links}
  \vspace{-5mm}
\end{figure}

\begin{table}[tb]
\caption{Four levels of cross-view data relationships.}
\label{4levels}
\resizebox{\columnwidth}{!}{%
\begin{tabular}{l|c|c}
\multicolumn{1}{c|}{\textbf{Level of Relationships}} & \textbf{Number of Views} & \textbf{Number of Visual Elements} \\ \hline
\textit{Individual} Level $(1 : 1)$            & 2          & 2               \\ \hline
\textit{Group} Level $(1 : i)$                 & 2          & $1 + i\ (i \geq 2)$             \\ \hline
\textit{Bi-group} Level $(i : j)$              & 2          & $i + j\ (i, j \geq 2)$           \\ \hline
\textit{Multi-group} Level $(i : j : ... : z)$ & at least 3 & $i + j + ... + z \ (i, j, ..., z \geq 2)$ 
\end{tabular}%
}
  \vspace{-4mm}
\end{table}


\subsection{Visual Element}
By following the definition of \textit{mark} and \textit{channel} discussed by Munzer \cite{munzner2014visualization}, we denote a visual element as \textit{a graphical representation unit} that encodes data.
The appearance of this unit can be controlled based on certain data attributes.
We consider that visual elements serve a key role in transforming data from their raw format into a human understandable format. 
This emphasizes a mapping from data to graphical representation units.
For example, nodes in a scatterplot correspond to data entities and the positions of nodes reveal values of two attributes.
Based on this data-mapping oriented notion, we do not consider user interface widgets (e.g., button, slider, and checkbox) as visual elements, although they are commonly used in visualizations.

\subsection{View and Multiple Views}
We define a view as a set of visual elements that are spatially organized in a perceivable visual-boundary to support specific analysis tasks.
A view can be a window of a desktop application (\eg, Jigsaw's List View \cite{stasko2008jigsaw}), a card in a dashboard visualization, a chart of a bounded area in a web application (e.g., a chart with visible borders in MyBrush \cite{koytek2018mybrush}), a component of a meta-visualization (\eg, a block in Domino \cite{gratzl2014domino}), or a chart of small multiples \cite{tufte1983visual}.
For some complex cases (\eg, a composite visualization \cite{javed2012exploring} in which one visualization is overlaid on top of another or embedded in another, like Interver \cite{sun2016interver}), we consider them one view, because they are intertwined with each other.
We treat multiple views as a technique that uses more than one view for data analysis, and these views are organized in a display space with certain layout strategies \cite{chen2020composition}. 

\subsection{Cross-View Data Relationship}

In terms of relationship components, there are three types of cross-view data relationships (Figure \ref{3links}). 
The first type highlights \textit{relationships between visual elements} from different views (e.g., three persons in a social network are related to four locations on a map). 
The second type emphasizes \textit{relationships between views}.
It may indicate high-level insights (e.g., placing documents near each other indicates they are relevant in ForceSPIRE \cite{endert2012semantic}).
The third type refers to \textit{relationships between visual elements and views}. 
It can be considered following Shneiderman’s visual information seeking mantra \cite{shneiderman2003eyes}.
For example, a line chart shows an overview of car sales and a bar chart displays detailed data (\eg, sales at different dealers) for a line in the overview.
In this work, we focus on the first type.
Compared to others, it requires users exploring detailed connections between visual elements.
Such low-level connections lay a necessary foundation for gaining high-level insights (\eg, matching between keywords leads to an understanding of relevant documents).
\rv{Based on the five levels of relationships} discussed in \cite{sun2014five}, there are four types of visual-element oriented cross-view data relationships. 
\rv{Table \ref{4levels} gives a summary of them, where $i, j, ... , z$ denotes the number of visual elements in each view.}

\section{Designing \name{}}



\subsection{High-Level User Tasks}
\label{sec-user-task}

We have analyzed user tasks when using MV, particularly focusing on the role of cross-view data relationships. 
We performed the analysis based on the dataset used in a prior study on composition and configuration patterns in MV \cite{chen2020composition}. 
It includes papers on MV in IEEE VIS, EuroVis, and PaciﬁcVis from 2011 to 2019.
For each paper in this dataset, we analyzed how multiple views were used and how cross-view data relationships supported such usage.
We have identified three types of user tasks:1) \textit{filtering}-oriented tasks, 2) \textit{refocusing}-oriented tasks, and 3) \textit{connecting}-oriented tasks.
They are supported by three different roles of cross-view data relationships, as summarized in Table \ref{tab:user-tasks}.

\textbf{Filtering-oriented tasks} refer to users selecting elements in one view, which filters elements in other views.
A typical example includes Cross-filtered views \cite{weaver2010cross} and cross-filtering based dashboard visualizations \cite{sarikaya2018we}, in which each view can be used to filter data on others. 
For such tasks, the view that users interact with serves a role similar to a control panel that enables users to perform dynamic queries \cite {ahlberg2003visual} on other views.
It implies a hierarchy between views (\eg, one controls others).
Filtering-oriented tasks are not a focus of our design, as the key information that users care about is filtered data in other views.
Thus, cross-view data relationships only provides a way that enables users to filter data, but the relationships are not a key analytical focus.

\begin{table}[tb]
\caption{Identified user tasks when using cross-view data relationships.}
\label{tab:user-tasks}
\resizebox{\columnwidth}{!}{%
\begin{tabular}{l|l}
\multicolumn{1}{c|}{\textbf{High-Level User Tasks}} & \multicolumn{1}{c}{\textbf{The Role of Cross-view Data Relationships}} \\ \hline
Filtering-oriented  & Enabling cross-view data filtering          \\ \hline
Refocusing-oriented & Offering one-to-one data mapping \\ \hline
Connecting-oriented & Serving analytical solutions      
\end{tabular}%
}
\vspace{-4mm}
\end{table}

\textbf{Refocusing-oriented tasks} are users selecting data in one view and based on the selection, users trying to explore the same data in other views.
As a simple example, a scatter plot matrix supports this type of task.
In a scatterplot matrix, each scatter plot presents the same set of data points with different attributes.
When users select some data points in one scatter plot, the same data points in other scatter plots are highlighted.
For refocusing-oriented tasks, cross-view data relationships offer a one-to-one data mapping that supports users in shifting their focus from one view to others.
Thus, 
key information that matters to users is how selected data in one view is presented in other views, rather than the relationship across views. 

\textbf{Connecting-oriented tasks} are users exploring and identifying connections between data displayed in multiple views.
Supporting this type of task is the key focus of our design for two reasons. 
First, cross-view data relationships are critical for this type of task.
The other two tasks care about views to be investigated next by using cross-view data relationships for making transitions from working on one view to others.
However, connecting-oriented tasks emphasize data relationships between views.
Second, as this type of tasks handles connections between data from different views and such connections may involve various combinations, 
it is more exploratory in nature than the other two tasks.
Solving a combinatorial problem manually is challenging and needs computational support (\eg, finding possible combinations, and showing them in a way that humans can understand).

\subsection{Design Trade-off}
\label{sec-trade-off}

\begin{figure}[tb]
  \centering
  \includegraphics[width=\columnwidth]{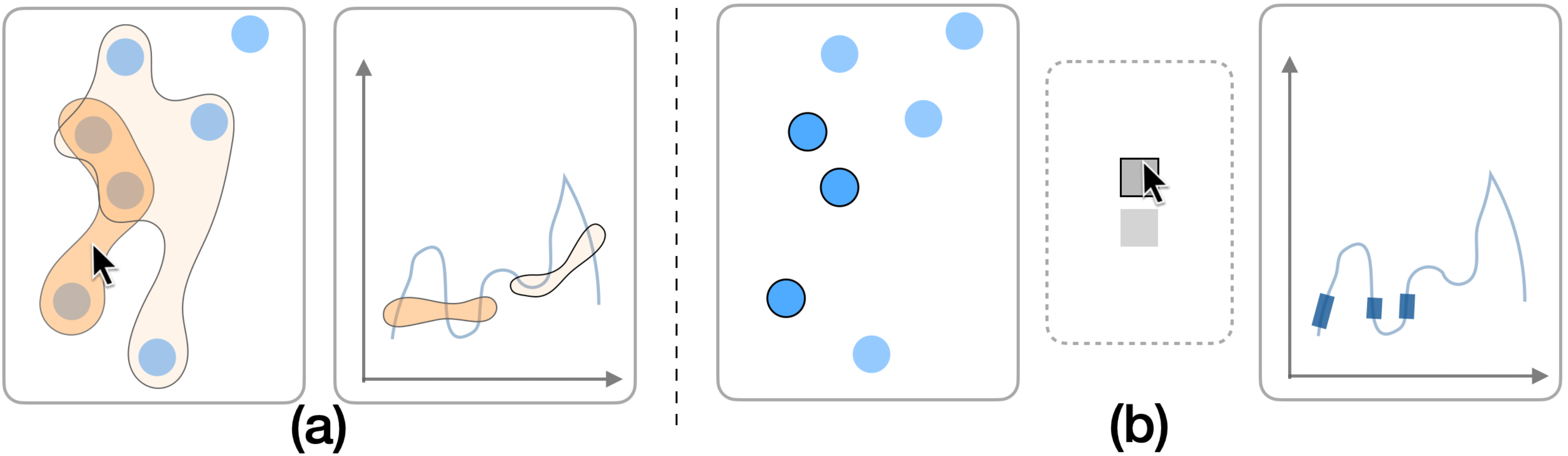}
    \vspace{-6mm}
  \caption{A trade-off in visualizing cross-view data relationships: (a) \textit{intertwined} with existing views, and (b) \textit{separated} from existing views.}
  ~\label{fig-trade-off}
    \vspace{-5mm}
\end{figure}

There is a key design trade-off in showing cross-view data relationships: \textit{view-embedding} (\textit{intertwined with} existing views) versus \textit{view-separating} (\textit{separated from} existing views), particularly when using new marks in addition to existing visual element marks for the relationship.
A comparison between them is summarized in Table \ref{tab-trade-off}.
Figure \ref{fig-trade-off} shows an example in which there are two bi-group level cross-view data relationships consisting of visual elements from two views: a scatterplot and a line chart.
Specifically, one is among five nodes and two line segments, and the other is among three nodes and three line segments.
They are revealed with marks indicating relationships either on top of the two views (Figure \ref{fig-trade-off} (a)) or outside of  the two views (Figure \ref{fig-trade-off} (b)).

\textbf{View-embedding designs} highlight encoding relationships inside existing views, such as using the BubbleSets \cite{collins2009bubble} technique to enclose visual elements of the same relationship.
A benefit of this is that a relationship is shown in the context of its involved visual elements, and it does not take extra space outside of existing views.
However, such embeddings force displayed relationships intertwined with existing visual elements, which can impact human perception of them.
For example, due to many relationship-marks overlaying each other, user attention to existing views may be directed towards the highly overlapped ones (as visually they look salient), whereas users may focus on other visual elements when no such view-embedding is present.

\textbf{View-separating designs} call for displaying relationships outside of existing views by using some empty display space, which can be considered creating a view for relationships.
A key benefit of this is that user perception of existing views is not disturbed. 
It allows users to find relationships in some specific areas in the display, which can facilitate the search for relationships.
However, to achieve such a separation, extra display space is necessary to hold the relationship-marks.
This requires that all existing views do not take up all the available display space. 
Moreover, with this separation, users need to switch their focus between existing views and the area where relationship-marks reside to check the detail of relationships.
Such a context-switching between visual elements and relationships entails extra cognitive effort \cite{convertino2003exploring}.

\begin{table}[tb]
\caption{A summary of key trade-offs between two designs.}
\label{tab-trade-off}
\resizebox{\columnwidth}{!}{%
\begin{tabular}{l|c|c}
                                                      & \multicolumn{1}{c|}{\textbf{View-Embedding}} & \multicolumn{1}{c}{\textbf{View-Separating}} \\ \hline
\multicolumn{1}{l|}{Relationship-marks placed}         & Inside existing views            & Outside existing views           \\ \hline
\multicolumn{1}{l|}{Empty space among views}    & Unnecessary                      & Necessary                      \\ \hline
\multicolumn{1}{l|}{Perception of existing views}    & Disturbed               & Preserved                \\ \hline
\multicolumn{1}{l|}{Get a relationship-overview} & Hard                    & Easy                     \\ \hline
\multicolumn{1}{l|}{Check relationship detail} & Less effort for context switching & More effort for context switching 
\end{tabular}%
}
    \vspace{-1mm}
\end{table}

\subsection{Design Considerations}

\rv{Based on the analysis discussed in previous sections and the knowledge tasks presented by Amar and Stasko in \cite{amar2004best}}, we have identified four major considerations that drive the design of \name.


\begin{table*}[!h]
\caption{A summary of six ways of user explorations with specific tasks.}
\label{tab-user-explorations}
\resizebox{\textwidth}{!}{%
\begin{tabular}{l|l|l|l}
\multirow{2}{*}{} &
  \multicolumn{1}{c|}{\multirow{2}{*}{\textbf{In the Same View}}} &
  \multicolumn{2}{c}{\textbf{Switching Views}} \\ \cline{3-4} 
 &
  \multicolumn{1}{c|}{} &
  \multicolumn{1}{c|}{\textbf{Switching to Existing Views}} &
  \multicolumn{1}{c}{\textbf{Switching to Relationship-Views}} \\ \hline
\multicolumn{1}{c|}{\textbf{\begin{tabular}[c]{@{}c@{}} Entity-\\ Driven\\ Exploration\end{tabular}}} &
  \begin{tabular}[c]{@{}l@{}}T1: Given a set of visual elements in an existing view,\\ find related visual elements in the same view. \\ \colorbox{hcolor}{(\textit{from entity to entity})} \end{tabular} &
  \begin{tabular}[c]{@{}l@{}}T2: Given a set of visual elements in an existing view, \\ find related visual elements in other existing views. \\ \colorbox{hcolor}{(\textit{from entity to entity})} \end{tabular} &
  \begin{tabular}[c]{@{}l@{}}T3: Given a set of visual elements in an existing view, \\ find related relationships in \rv{relationship-views}. \\ \colorbox{hcolor}{(\textit{from entity to relationship})} \end{tabular} \\ \hline
\multicolumn{1}{c|}{\textbf{\begin{tabular}[c]{@{}c@{}} Relationship-\\ Driven \\ Exploration\end{tabular}}} &
  \begin{tabular}[c]{@{}l@{}}T4: Given a set of relationships in a \rv{relationship-view},\\ find related relationships in the same view. \\ \colorbox{hcolor}{(\textit{from relationship to relationship})} \end{tabular} &
  \begin{tabular}[c]{@{}l@{}}T5: Given a set of relationships in a \rv{relationship-view},\\ find related entities in existing views. \\ \colorbox{hcolor}{(\textit{from relationship to entity})} \end{tabular} &
  \begin{tabular}[c]{@{}l@{}}T6: Given a set of relationships in a \rv{relationship-view}, \\ find related relationships in other \rv{relationship-views}. \\ \colorbox{hcolor}{(\textit{from relationship to relationship})}\end{tabular} 
\end{tabular}%
}
\vspace{-4mm}
\end{table*}

\textbf{C1: Designing relationship encodings}. 
Effective encodings for relationships are critical, which should fulfill four goals, discussed below.
\rv{They correspond to the knowledge task, concretizing relationships \cite{amar2004best}, which calls for clearly showing detailed components of relationships.}

(a) Relationship encodings should be perceptually discriminated from the encodings of visual marks in existing views, so that users can easily recognize relationships. 
(b) Relationship encodings should be designed to avoid disturbing user perception of existing views.
This helps preserve existing views.
(c) Relationship encodings should reveal visual hints about visual elements that belong to the relationship to guide user exploration.
This helps users check detailed information of a relationship.
(d) Visual encodings for relationships should reveal changes when the state of a relationship is updated (\eg, selected vs. unselected).
(a)-(c) calls for a balanced-solution regarding the design trade-off discussed in Section \ref{sec-trade-off}.


\textbf{C2: Supporting six ways of explorations}.
For connecting-oriented tasks, there are six possible ways of user explorations, as summarized in Table \ref{tab-user-explorations}. 
We aim to support all to enable flexible user explorations.
User explorations can be considered in two dimensions.
One cares about user explorations driven by \textit{entity} (\eg, visual elements in existing views) or \textit{relationship}. 
The other highlights whether user explorations remain in the \textit{same view} or \textit{switch views}.
Cross-view data relationships involves visual elements (as entities) and connections between entities (as relationships). 
Users can start explorations with either of them. 
This leads to four different explorations: from entity to entity, from entity to relationship, from relationship to entity, and from relationship to relationship. 
As entities are located in existing views and relationships can be shown outside of existing views (in some empty display area, which may form views for relationships), as discussed in Section \ref{sec-trade-off}, these four explorations can be further categorized into two major groups: in the same view (T1 and T4) or switching views.
Regarding switching views, there are two possible cases: \textit{switching to existing views} (T2 and T5) and \textit{switching to relationship-views} (T3 and T6).



\textbf{C3: Organizing relationships and views}. 
Relationship should be visually organized.
This can ease the search process for finding useful relationships.
Moreover, an organized layout may serve as an overview of relationships, which can better guide user explorations than a layout with relationships randomly placed. 
Furthermore, users may want to flexibly organize multiple views in a personalized, meaningful way for sensemaking (\eg, using spatializations) \cite{andrews2010space}.

\textbf{C4: Referring to original data on demand}.
To help users check and understand computed cross-view data relationships, showing original data is necessary.
Users need to refer to original data (e.g., documents) to learn detailed context, which helps them verify computed relationships that may imply certain hypotheses.
\rv{This corresponds to the knowledge task, confirming hypothesis, discussed in \cite{amar2004best}.}
Moreover, original data should be retrieved on demand (e.g., requested based on some relationships, instead of simply showing all data).


\section{\name\ Technique}
The \name\ technique comprises three parts: a) relationship computation, b) relationship visualization based on a \textit{bi-context} design, and c) interactive relationship management.

\subsection{Cross-View Data Relationship Computation}
We focus on supporting connecting-oriented tasks, as discussed in Section \ref{sec-user-task}, which require users to explore sets of related visual elements from different views.
As biclustering is designed for mining related subsets of entities from two entity-sets, 
we can formalize cross-view data relationships as biclusters. 
However, how to compute such biclusters remains a problem.

\subsubsection{Challenges for Computation}
\label{compute-challenge}

\begin{figure}[tb]
  \centering
  \includegraphics[width=\columnwidth]{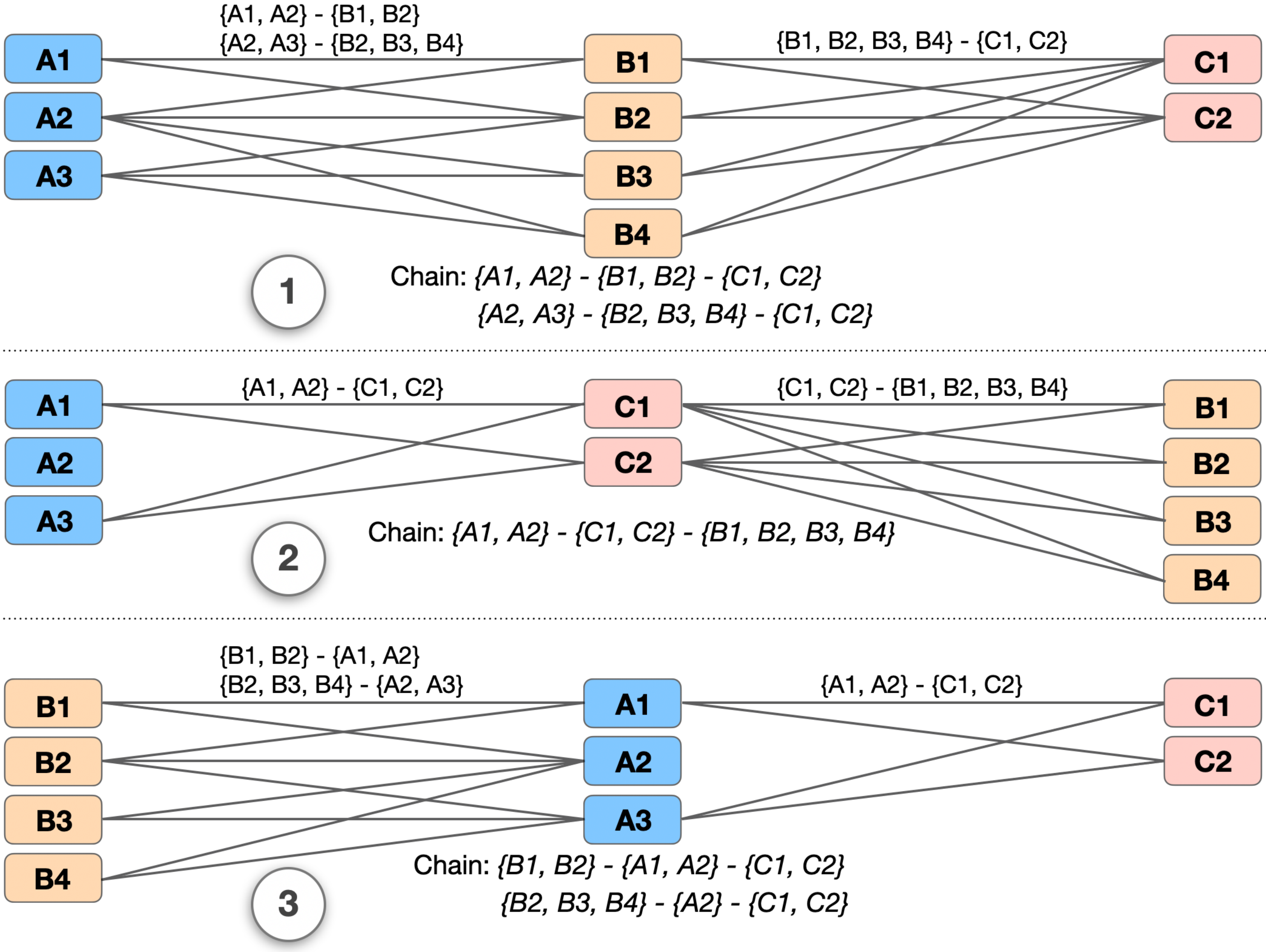}
    \vspace{-6mm}
  \caption{For the same sets of entities, different combinations of biclusters by permuting the order of the three sets lead to different bicluster-chains.}
  ~\label{fig-bic-chain-view}
    \vspace{-5mm}
\end{figure} 

There are two challenges in computing cross-view data relationships.
First, since we use biclusters to formalize cross-view data relationships, to enable computing biclusters consisting of visual elements from different views, we need support from a certain \textit{data model}.
Second, for multi-group level of relationships, while they can be further formalized as \textit{bicluster-chains} (as more views are involved), computing them is not simple.
Given a set of views, there are multiple ways of chaining biclusters, as the sequence of views can vary (see Figure \ref{fig-bic-chain-view}).
For the same three sets of entities, using different sequences to connect them leads to different bicluster-chains, and entities in one bicluster-chain can be subsets of those in another (Figure \ref{fig-bic-chain-view} \circled{1} and \circled{2}).

\subsubsection{Data Model for Computing Biclusters}
\label{sec-data-model}


To address the challenges, \name\ uses a relational data model to compute biclusters between visual elements from a pair of views.
With this model, we can get biclusters that reveal cross-view data relationships. 
It lays the foundation to fulfill the four design considerations.
Specifically, this model can be created with three key steps, shown in Figure \ref{fig-data-model}.

\textbf{S1:} We tabulate visual elements by generating a table for each existing view.
In this table, each visual element has a unique ID, and the data that this visual element encodes are associated with the ID.
This step highlights a reverse mapping between data and marks in each view, and further organizes them in a table.

\textbf{S2:} For all pairs of views, based on related data encoded by visual elements, we create a joint table that associates the IDs of visual elements from one view with those from another.
The way that determines how data is related can be different for different scenarios.
\rv{For example, in text analytics, two people and three locations are related based on word co-occurrence; whereas in cyber security, five MAC addresses are related to four URLs determined by network traffic patterns \cite{zhang2015visualizing}.}

\textbf{S3:} We transform each joint table into a data matrix, where rows and columns are visual element IDs from two views.
For a cell with a corresponding row and column ID pair in the joint table, we assign a value of 1, otherwise 0.
For each data matrix generated in this step, we can apply biclustering algorithms (e.g., LCM \cite{uno2004efficient} and CHARM \cite{zaki2005efficient}) to it.
Such computed biclusters can reveal cross-view data relationships, particularly between a pair of views.



\subsubsection{Bicluster Chain Computation}

As discussed in Section \ref{bic}, we can form bicluster-chains based on shared entities.
This allows us to expand cross-view data relationships to cover visual elements from more than two views (\eg, a set of nodes from a network graph, a set of line segments from a line chart, and a set of locations on a map).
A key benefit of this approach is that it starts with a pair of views and supports users in flexibly including more views to expand explorations of relationships as their analysis proceeds.
This dynamic expansion further enables users to interactively organize computed relationships.
To address the second challenge of various ways of chaining biclusters, \name\ uses a four-step approach to get connected sets of visual elements from each involved view:

\begin{figure}[tb]
  \centering
  \includegraphics[width=\columnwidth]{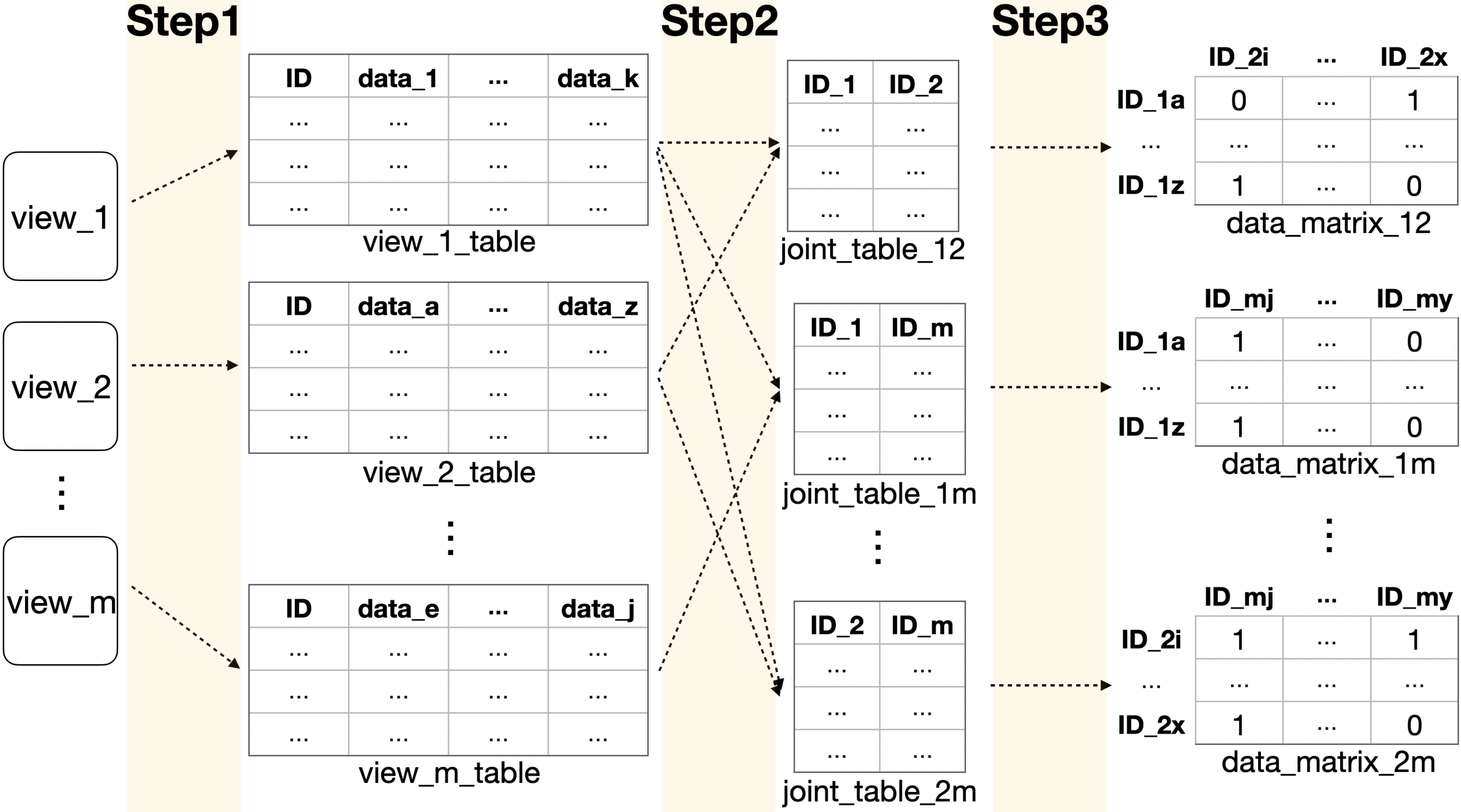}
    \vspace{-6mm}
  \caption{Step1: reverse mapping from visual elements to a table for each view. Step2: creating a joint table for each pair of tables from previous step. Step3: transforming each joint table into a data matrix.}
  ~\label{fig-data-model}
    \vspace{-5mm}
\end{figure} 


\textbf{S1:} \textit{Computing biclusters for all pairs of views.}
For a given set of views (\eg, \{$A$, $B$, $C$\}), we get all pairs of them (\eg, \{$AB$, $BC$, $AC$\}), and for each pair, we compute biclusters composed of visual elements from this pair of views.

\textbf{S2:} \textit{Getting sequences for a given set of views.}
For the same given set of views in the previous step, we compute their permutation without reverse duplicates (\eg, \{$ABC$, $ACB$, $BAC$\}).
We do not include reverse duplicates, because bicluster-chains for a sequence (\eg, $ABC$) and its reverse duplicate (\eg, $CBA$) are formed by the same two sets of biclusters (\eg, biclusters computed based on $AB$ and $BA$ are the same).

\textbf{S3:} \textit{Building bicluster-chains.}
Based on the sequences from S2, we separate each sequence of the views (e.g., $ABC$) into consecutive, neighboring pairs (\eg, $AB$ and $BC$).
For each neighboring pair, we find computed biclusters from S1.
Following the sequence of neighboring pairs, we chain corresponding biclusters together based on their shared entities in the same view (\eg, view $B$).
Specifically, we use equation (\ref{equ-matching}) to compute the matching between two biclusters that share visual elements in one view. 

Given two biclusters consisting of visual elements from view pairs $AB$ and $BC$, $bic_{AB} = \{\{a_i\}, \{b_j\}\}$ and $bic_{BC} = \{\{b_m\}, \{c_k\}\}$, where $a_i \in A$, $b_j, b_m \in B$, and $c_k \in C$, and $i, j, m, k \geq 1$, the matching between them is computed as follows ($|\cdot|$ denotes the cardinality of a set):

\begin{equation}
\label{equ-matching}
matching(bic_{AB},\ bic_{BC})= \frac{|b_j \cap b_m|}{|b_j \cup b_m|}
\end{equation}

Based on the computed matching score, we determine whether two biclusters are chained or not (\eg, chaining them if the matching score between them is above a threshold).
With this step, we get all possible bicluster-chains for the given set of views.

\begin{algorithm}[tb]
    \SetKwInOut{Input}{Input}
    \SetKwInOut{Output}{Output}

    \Input{\textit{allBicChains}, a set of all chains $\{c_i, i = 1\ ...\ n\}$\newline 
    \textit{entChainDict}, a dictionary with elements for each chain}
    \Output{\textit{selBicChains}, a set of selected chains}
        
    \caption{Cleaning obtained chains to get selected ones}
    
    \For{$i \leftarrow 1\ ...\ n$}{
    		$entSet_i \leftarrow$ \textit{entChainDict}($c_i$)\;
		 \For{$j \leftarrow i + 1\ ... \ n$}  {
		 	$entSet_j \leftarrow$ \textit{entChainDict}($c_j$)\;
		 	\If{ $entSet_i \not\subseteq entSet_j$ \textup{and} $entSet_j \not\subseteq entSet_i$}{
				\textit{selBicChains}.add($c_i$)\;
   			}
		 }
	}
	\Return \textit{selBicChains}\;
\end{algorithm}

\begin{figure*}[tb]
  \centering
  \includegraphics[width=\textwidth]{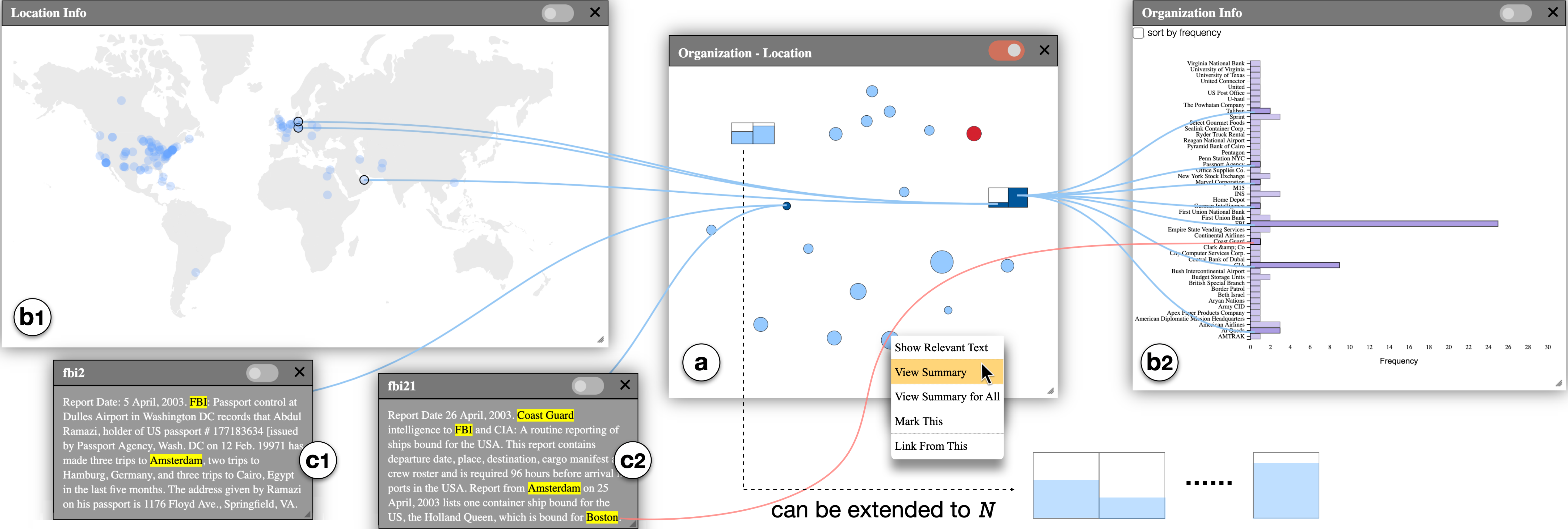}
    \vspace{-6mm}
  \caption{Visual encodings in \name. 
  (a) is a relationship-view showing computed relationships between locations on a map (b1) and organizations in a list (b2). 
  (c1) and (c2) display text, from which locations and organizations have been abstracted. 
  In (a), each relationship is shown as a circle and its involved visual elements in (b1) and (b2) can be linked via curved lines.
  The radius of a circle reveals the number of elements in this relationship and the relative position between circles indicates the similarity between two relationships.
  Using a right-click menu on each circle, users can create visual links between visual elements across different views and view the relationship in detail, which visually transforms a circle into a mini bar chart.}
  ~\label{fig-encoding}
    \vspace{-7mm}
\end{figure*} 

\textbf{S4:} \textit{Getting selected bicluster-chains.}
Based on the set of bicluster-chains from S3, for each chain, we check two cases to decide whether to keep it as a selected one.
One case is whether this chain is a subset of any other chain(s) in the set. 
The other is whether any other chain(s) in the set is a subset of this chain.
If neither is true, we consider this chain a selected one.
Algorithm 1 shows how to check whether a chain is a subset of another.
We remove the chain that is a subset of another, since it does not fully cover related visual elements from the given views.
This step ensures that there is no inclusion between bicluster-chains.

\subsection{Cross-View Relationship Visualization}
\name\ uses a \textit{bi-context} design. 
It separates visual marks that encode computed cross-view data relationships from existing views (\textbf{C3}).

\subsubsection{Bi-Context Design Concept}
We propose a bi-context design concept to visualize cross-view data relationships in \name.
It highlights that enriching MV by creating extra views to show computed data relationships across different views.
We refer to this newly created view as a \textit{relationship-view} to differentiate it from existing views.
\name\ allows users to create relationship-views for two-level relationships: 1) \textit{bi-group level}, and 2) \textit{multi-group level}.
The former is based on biclusters computed between visual elements from pairs of existing views. 
The latter is based on bicluster-chains that involve visual elements from more than two existing views.

Relationship-views enrich MV since they transform empty display space into views with useful data.
Augmented with relationship-views, MV can offer two types of context.
One consists of existing views. 
The other comes from newly added relationship-views.
As they are separated, it helps users ditinguish computed relationships from their involved entities (\textbf{C1-a}).
Also, this separation helps preserve existing views, so user perception of them is not significantly affected by newly added marks for relationships (\textbf{C1-b}).
This bi-context design supports all six ways of user explorations (\textbf{C2}).
With newly created relationship-views, users can choose where to start explorations (\ie, an existing view or a relationship-view) and switch the context usage. 


\subsubsection{Visual Encodings for Cross-View Data Relationships}
Based on computed biclusters and bicluster-chains, discussed in Section \ref{sec-data-model}, \name\ supports creating relationship-views for bi-group level and multi-group level of cross-view data relationships.
\name\ uses the same visual encodings for them.
Figure \ref{fig-encoding} shows an example of detailed visual encodings of a bi-group level relationship-view (a), which relates locations in a map (b1) with organizations in a list (b2).

A relationship-view shares the same UI components with existing views, such as a panel header, a panel body, a pin button (\scalerel*{\includegraphics{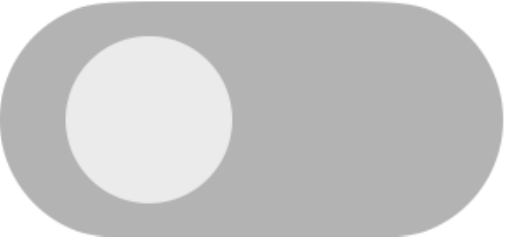}}{B}), and a close button (\scalerel*{\includegraphics{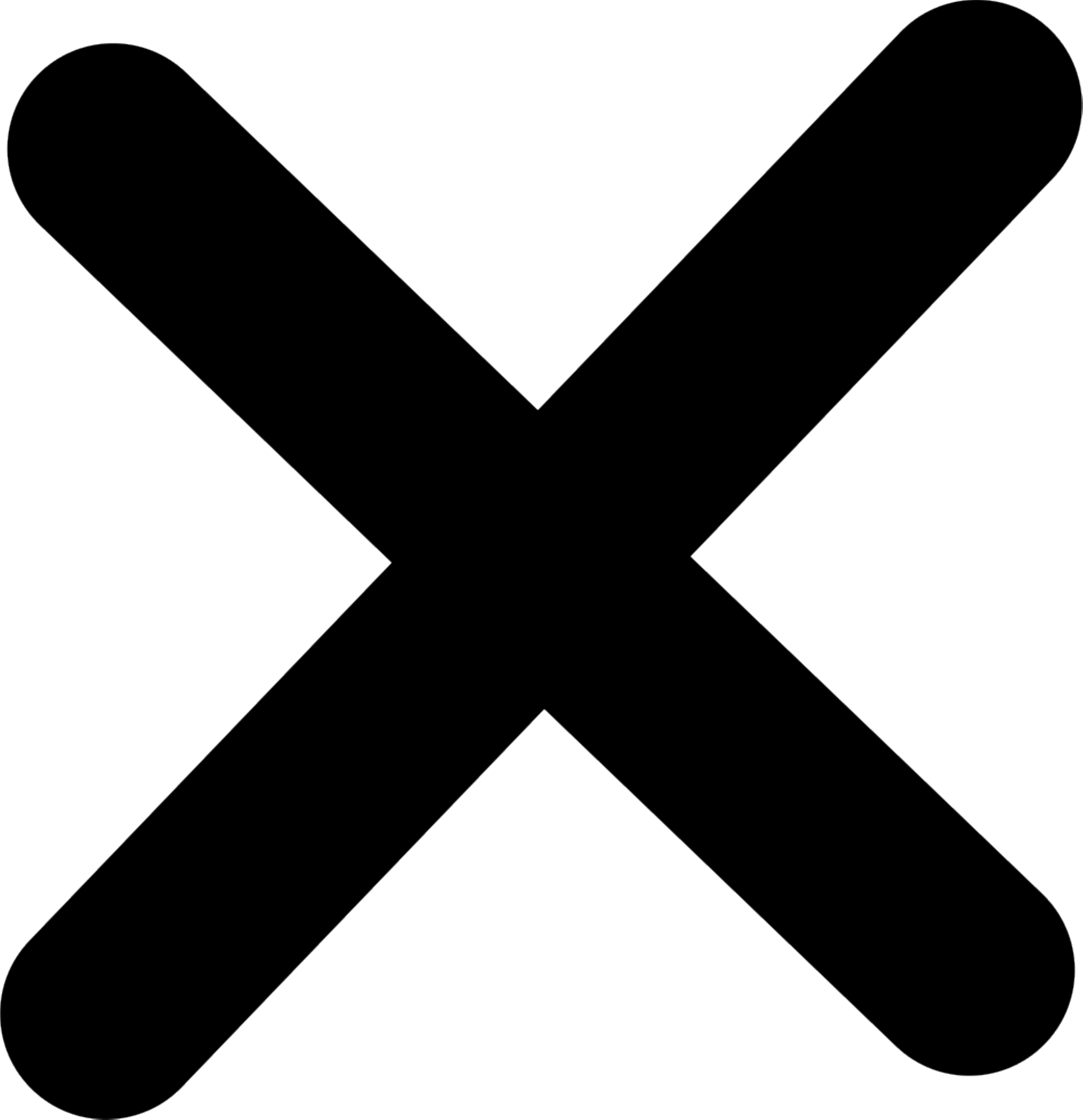}}{B}).
In the body of a relationship-view, computed biclusters or bicluster-chains are displayed.
Each bicluster or bicluster-chain is displayed as a circle (\scalerel*{\includegraphics{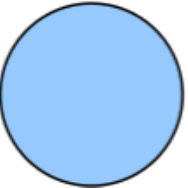}}{B}).
Its radius reveals the total number of involved elements by using a linear mapping function.
The more elements are in a bicluster or bicluster-chain, the larger radius a circle has.
Moreover, when users click or hover a mouse pointer on a circle, it gets darker in color (\includegraphics[height=\fontcharht\font`\B]{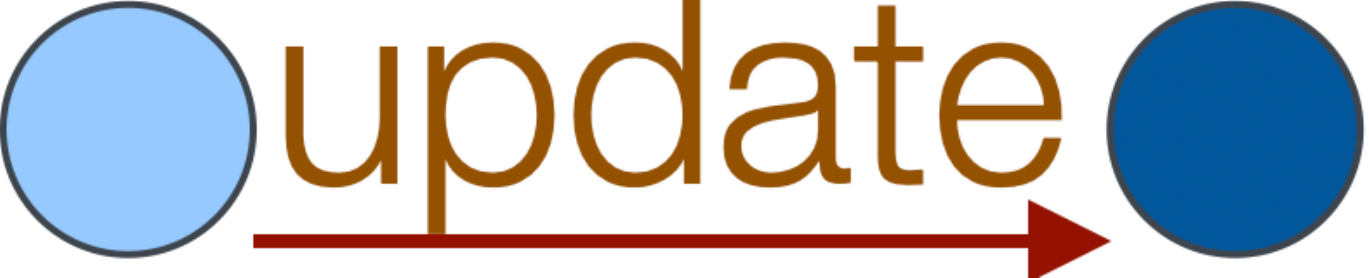}).
Also, the color of a circle changes to red (\includegraphics[height=\fontcharht\font`\B]{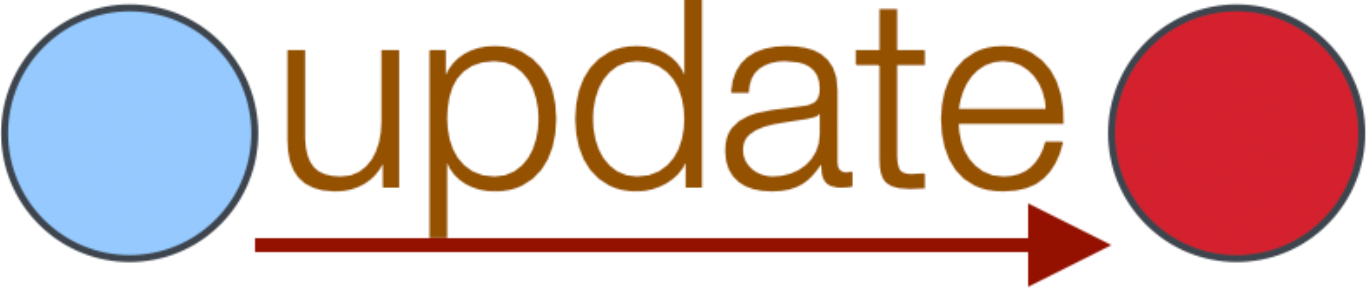}), after a user chooses to mark it by using a right-click menu on the circle.
Such color updates reveal that the state of a bicluster or bicluster-chain has changed from normal to selected, focused or marked (\textbf{C1-d}).

The positions of circles are computed with multidimensional scaling (MDS) \cite{cox2008multidimensional}, so the relative distance between two circles reveals the similarity between two biclusters or bicluster-chains (Table~\ref{tab-user-explorations}, \textbf{T4}). 
The closer two circles locate, the more similar two biclusters or bicluster-chains are. 
To apply MDS, we transform each bicluster or bicluster-chain into a vector. 
Each dimension of the vector corresponds to a visual element from a view that belongs to this bicluster or bicluster-chain.
With this transformation, we generate a $m \times n$ matrix, where $m$ is the number of biclusters or bicluster-chains, and $n$ corresponds to all visual elements in them.
The value of cells in this matrix is assigned as $1$, if a bicluster or bicluster-chain has a visual element; otherwise, it is $0$.
With this matrix, we compute the pairwise distance between biclusters or bicluster-chains, and then form a distance matrix in which both rows and columns are biclusters or bicluster-chains.
Such a distance matrix is then used as the input for MDS, which consequently generates the coordinates for each bicluster or bicluster-chain. 

For each bicluster or bicluster-chain, \name\ allows users to check it in detail (\textbf{C1-c}).
\name\ offers two levels of detail for a bicluster or bicluster-chain: a \textit{concrete} level and a \textit{summary} level.
A concrete level of detail allows users to see exact visual elements that belong to a bicluster or bicluster-chain, as visually linking a visual element with a bicluster or bicluster-chain via a curved line. (\includegraphics[height=\fontcharht\font`\B]{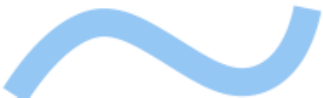} or \includegraphics[height=\fontcharht\font`\B]{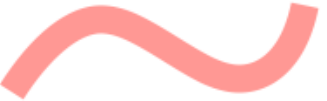}).
A blue curved line reveals an automatic visual linking. 
A red one is a user-created visual linking. 
They are discussed in detail in Section \ref{sec-interaction}.
By following the lines, a user can check the detail of a bicluster or bicluster-chain.
\name\ uses this by default, so such visual connections appear when a user selects or hovers on a circle in a relationship-view.
 
 
A summary level of detail offers users a quick overview of a bicluster or bicluster-chain. 
It changes a circle to a mini bar chart (\includegraphics[height=\fontcharht\font`\B]{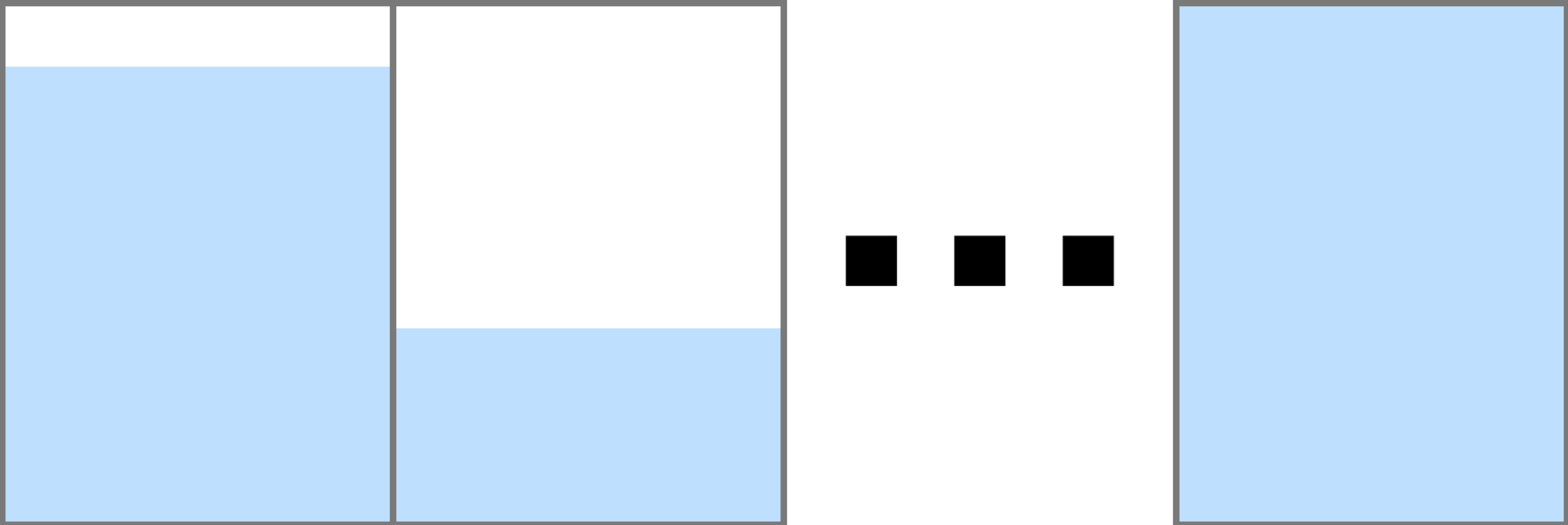}), where all bars are horizontally aligned.
The height of a bar reveals the number of visual elements in a view that belong to a bicluster or bicluster-chain.
The bars are ordered by the sequence of existing views shown in the display space.
For example, in Figure \ref{fig-encoding}, the left bar in a mini bar chart shown in (a) corresponds to the visual elements on the map (b1), as the map was added to the display space earlier than the list (b2).
Moreover, \name\ applies a bounding box, with the same size, in each bar.
They offer a visual reference to help users compare the number of involved visual elements from different views to gain a quick overview of how different views contribute to a bicluster or bicluster-chain.
Using a right-click menu on a circle, users can see its summary or choose to view the summary for all relationships.
After a user chooses this, the circle changes into a mini bar chart (\includegraphics[height=\fontcharht\font`\B]{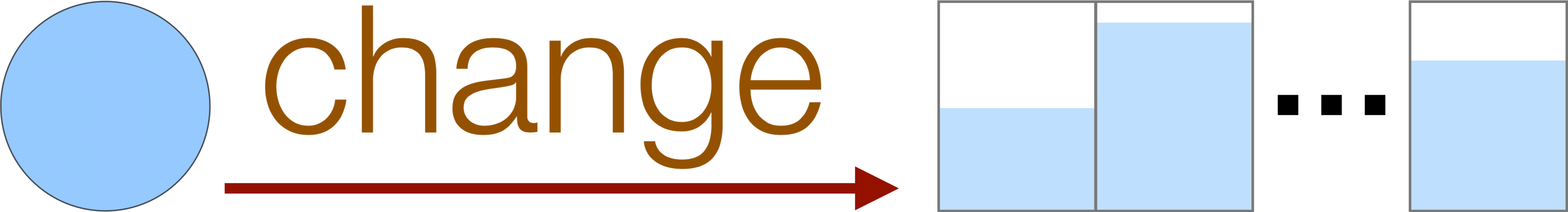}),
 and curved lines connected to the center of the circle, if displayed (\includegraphics[height=\fontcharht\font`\B]{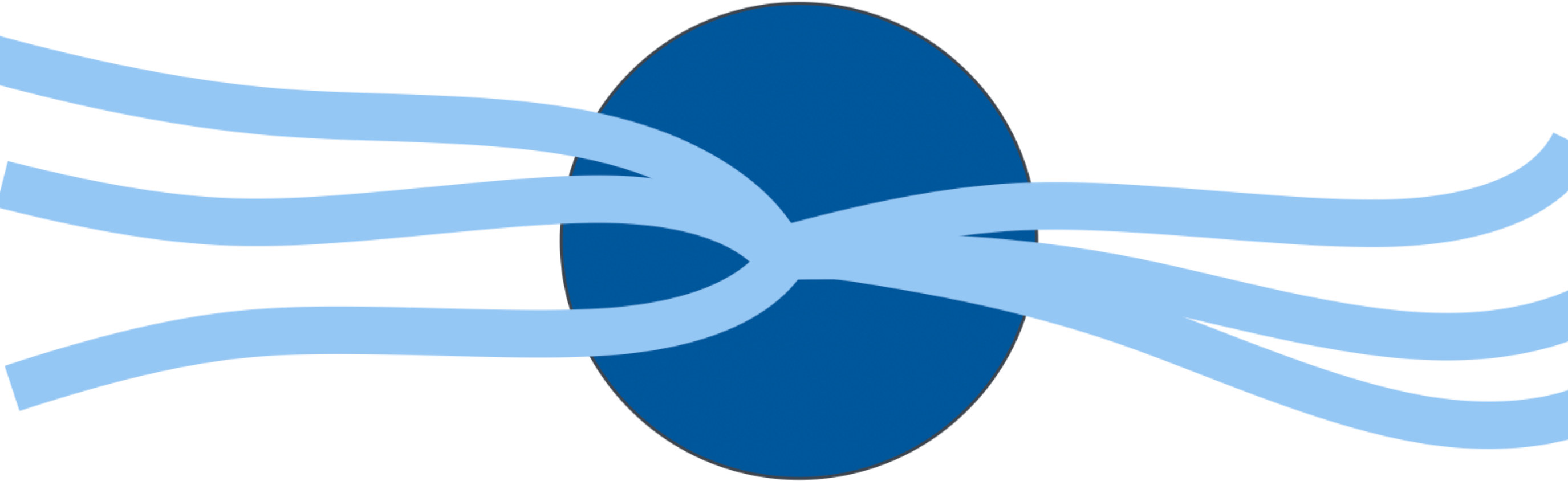}), switch to being linked to the center of different bars based on the view in which the connected visual elements are located (\includegraphics[height=\fontcharht\font`\B]{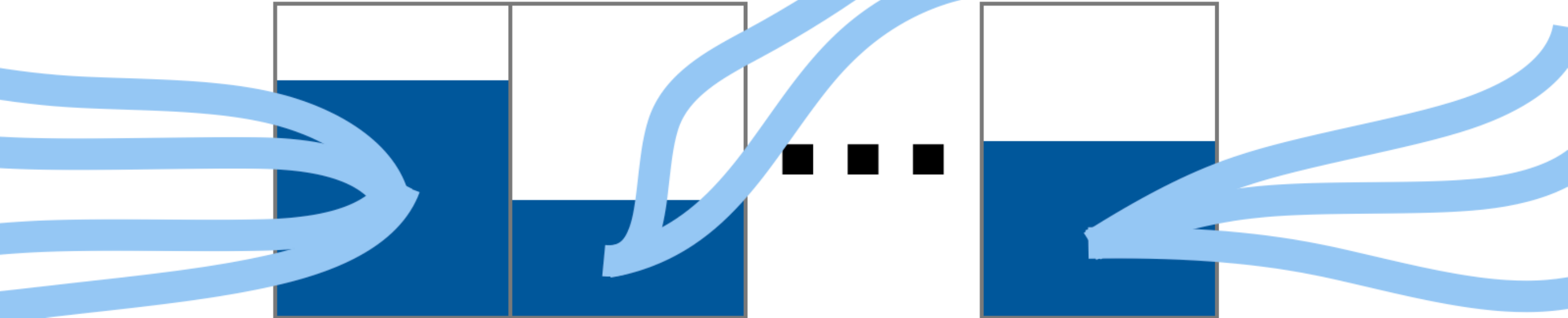}).

\subsection{Interactive Cross-View Relationship Management}
\label{sec-interaction}
\name\ offers a set of user interactions for using computed relationships. 
They serve as complements of the visual encodings, which are designed to help users view and retrieve related information (\textbf{C2} and \textbf{C4}) and organize information (\textbf{C3}).

\textbf{Resizing}. In \name, a user can change the size of a view by dragging a handler (\includegraphics[height=\fontcharht\font`\B]{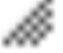}) on the bottom right corner of each view.
Enabling view resizing helps users get empty space, which leaves the room for creating relationship-views (\textbf{C1-a, b}).

\textbf{Pinning and Dragging}. 
\name\ allows flexible layout management at two levels (\textbf{C3}): \textit{view}-level (for organizing multiple views) and \textit{relationship}-level (for organizing relationship-marks in a relationship-view).
They are supported by enabling users to pin and drag a view, or drag relationship-marks in a relationship-view.
When pinning is disabled (\scalerel*{\includegraphics{figures/switch.pdf}}{B}) for a view, users can move it in the display space by dragging its header.
This allows users to create meaningful spatialization that may benefit from using space to think \cite{andrews2010space}.
When pinning is enabled (\scalerel*{\includegraphics{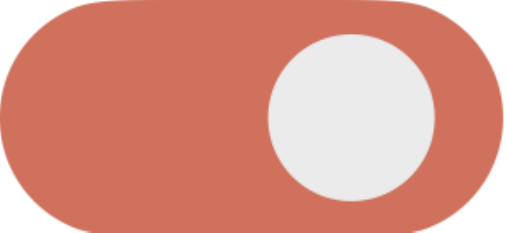}}{B}) for a view, it is undraggable. 
Moreover, \name\ applies the dust and magnet visual metaphor \cite{soo2005dust} to support users in moving several related views together. 
Specifically, as a user drags a relationship-view, other views in which some visual elements are visually connected to those in the dragged one via curved lines, can automatically move with the dragged one, if their pinning is disabled (see Figure \ref{fig-teaser}).
This helps users quickly pull multiple related views out of a cluttered spatialization (\textbf{C3}), such as some related views are covered by a few unrelated ones.
In addition to the view-level arrangement, for a relationship-view, if it is pinned, \name\ allows users to adjust the layout of displayed relationships by dragging them (\textbf{C3}).
This enables users to revise the layout generated based on MDS, which can reveal user understanding of the similarity of computed cross-view data relationships. 

\textbf{Visual linking}. 
\name\ supports two types of visual linking: \textit{automatic} linking and \textit{manual} linking.
They are revealed as blue/red curved lines
, respectively, connecting visual elements from different views and highlighting the connected visual elements.

Automatic linking shows such visual effects when a user selects or hovers on a visual element in a view or a relationship-mark in a relationship-view.
The connections between visual elements are found automatically based on the generated data matrix, computed biclusters or bicluster-chains, discussed in Section \ref{sec-data-model}.
The automatic linking aims to support six ways of user explorations (\textbf{C2}), by using a four-way search to create visual links.
The search starts with either visual elements in a view, or relationships in a relationship-view, which users interact with, and then checks visual elements in the same view (Table~\ref{tab-user-explorations}, \textbf{T1}) or different views (\textbf{T2}, \textbf{T5}), or relationships in the same relationship-view (\textbf{T4}) or different relationship-views (\textbf{T3}, \textbf{T6}).

 For a manual linking, \name\ allows users to specify visual elements to be linked.
 Specifically, users can flexibly create a visual link between visual marks in any views by using a right-click menu.
It works as a complement to the automatic linking, especially for connections based on domain knowledge but not explicitly mentioned in data. 


\textbf{Retrieving original data}.
\name\ provides a right-click menu on visual elements in a view, or relationship-marks in a relationship-view.
From the menu, users can choose to show related parts of original data (\eg, relevant text), which supports on-demand data retrieval (\textbf{C4}).
Retrieved data is shown in a stand-alone view (Figure \ref{fig-encoding}(c1) and (c2)).


\section{Usage Scenario}
\label{scenario}

\begin{figure*}[tb]
  \centering
  \includegraphics[width=\textwidth]{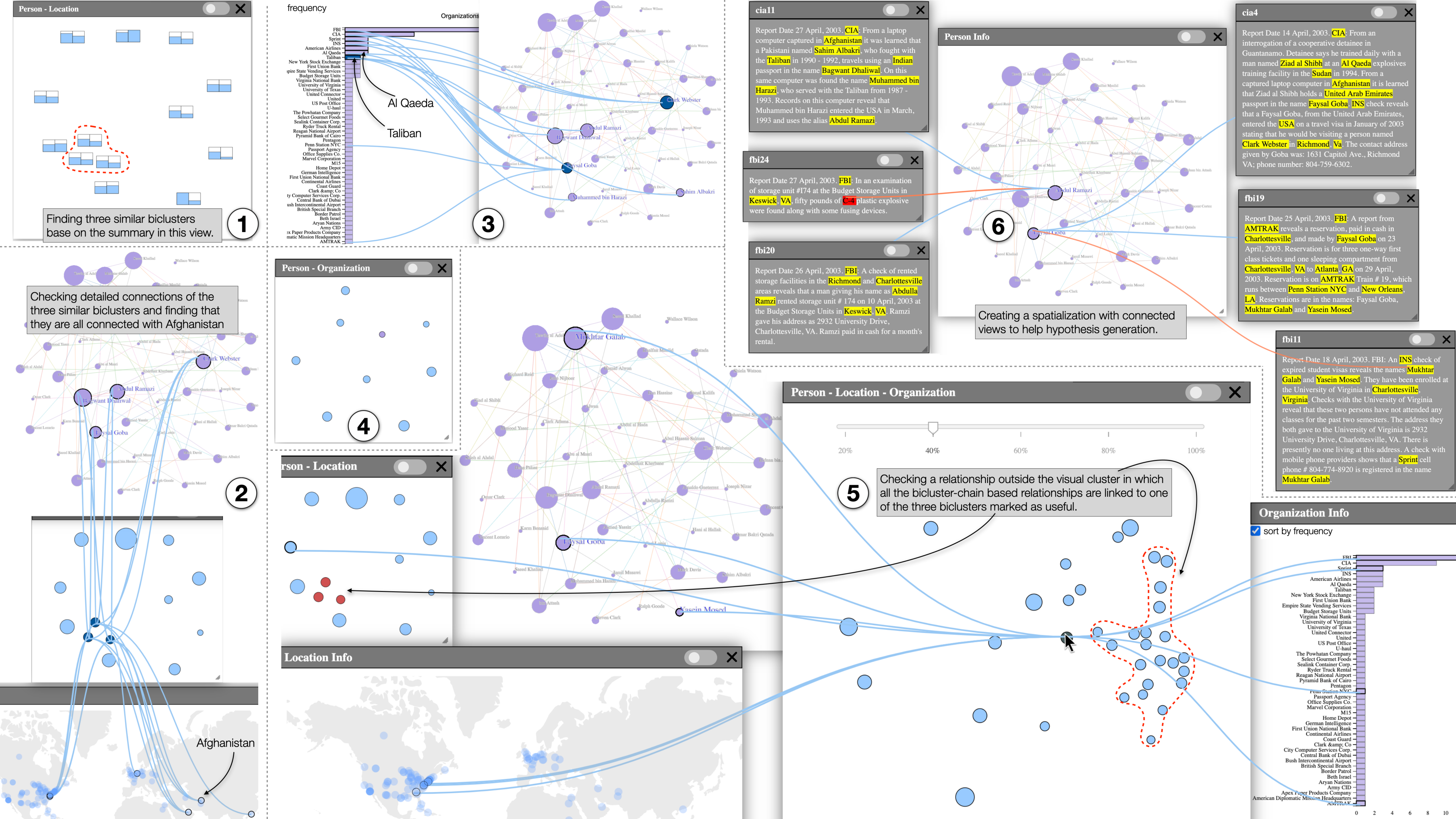}
    \vspace{-6mm}
  \caption{Key analytical steps of using \name\ to identify an important, colluded threat from the Sign of the Crescent dataset \cite{hughes2003discovery}.}
  ~\label{fig-case}
    \vspace{-7mm}
\end{figure*} 

We present a text analytics scenario that illustrates using \name\ to explore data relationships across multiple views to solve an investigative analysis problem.
The scenario explores the Sign of the Crescent dataset \cite{hughes2003discovery}, which has 41 fictional intelligence reports of suspicious activity.
It has been studied to understand human sensemaking process with real-world users in both traditional laboratory settings \cite{sun2014role}, and crowdsourcing settings \cite{li2018crowdia} using Amazon Mechanical Turk \cite{mturk}.

We extracted named entities (\eg, person, location, organization, etc.) from the text with NLTK \cite{nltk}.
For pairs of sets of named entities (\eg, person and location, person and organization, location and organization, etc.), based on word co-occurrence, we generated a relationship matrix in which each row is an entity from one set, each column is an entity from the other, and each cell is set to 1 or 0 depending on whether the two corresponding entities are related or not.
We applied LCM \cite{uno2004efficient} to each relationship matrix to compute biclusters.
In total, we had 296 named entities, 2404 entity-connections, and 158 biclusters.

Suppose that Ella is an investigator, who is given a task of identifying potential threats and key players from a collection of intelligence reports.
She loads the data into \name\ and starts her investigation.
Figure \ref{fig-case} illustrates key steps in Ella's analysis process.

\textbf{Overviewing data relationships between a pair of views.}
Ella starts analysis with two views, a graph view and a map view, in the \name\ workspace.
The former shows connections among persons in a graph. 
The latter displays locations mentioned in the reports on a map.
Glancing at them, Ella feels that the graph has too many edges and lots of locations have been reported. 
She thinks that checking each individually will not help solve the mystery.
Ella decides to explore relationships between them to see if any collusions exist among persons based on their involved events reported at the same locations.
As requested, \name\ shows a relationship-view between the two views in Ella's workspace. 
By looking at it, Ella quickly gets an overview of the relationships and finds that five circles are near each other, which seems similar (T4).
To verify this, she chooses to see the summary of all relationships in this relationship-view (Figure \ref{fig-case} \circled{1}), which changes circles to mini bar charts.
By checking the summary, Ella finds that three of the five neighboring clusters are similar (T4), as the number of involved elements from the two views are almost the same. 

\textbf{Switching the usage of bi-context}.
Following this finding, Ella changes the relationship-view to the simple mode that uses circles and checks the three similar ones in detail (Figure \ref{fig-case} \circled{2}), by switching her focus back to the graph view and the map view.
Ella hovers her mouse pointer on each circle and \name\ shows its detailed connections with elements in the graph and the map (T5).
By following the lines, Ella finds that they each connect with three of four frequently mentioned persons in the graph (circles of a relatively bigger size), and all are linked to \texttt{Afghanistan}, a sensitive location for Ella. 
Ella goes back to the relationship-view and marks the three investigated relationships as useful. 
Then, she decides to follow this lead to check more information, so Ella adds a list view of organization information to her workspace and sorts it by frequency.
In this list, two top-ranked and terrorist-related organizations, \texttt{Taliban} and \texttt{Al Qaeda} quickly catch her attention, so Ella decides to check connections between persons and them. 
As she hovers the mouse pointer on them, \name\ shows the linked persons (T2).
Ella finds that of the four previously noticed persons, two (\texttt{Abdul Ramazi} and \texttt{Bagwant Dhaliwal}) are linked with \texttt{Taliban}, and others (\texttt{Faysal Goba} and \texttt{Clark Webster}) are connected to \texttt{Al Qaeda} (Figure \ref{fig-case} \circled{3}).
After learning this, Ella wants to check if there are any similar clusters that group persons based on the two organizations (T3), so she adds another relationship-view, between the graph and the list, to her workspace and switches the focus of her analysis.
However, after seeing this relationship view (Figure \ref{fig-case} \circled{4}), Ella realizes that there are no obviously similar clusters because the circles are not near each other.

\textbf{Involving several views to gain deeper insights.}
As there are three types of information (persons, locations and organizations) displayed in three different views, Ella thinks that fusing the information together may lead to deeper insights.
Following Ella's request, \name\ includes all three views to compute cross-view data relationships, and adds a chain-level relationship-view to her workspace (Figure \ref{fig-case} \circled{5}).
Ella adjusts the threshold for computing bicluster-chains to $40\%$ and starts her examination.
By checking detailed connections between this chain-level relationship-view and the relationship-view between the graph (person) and the map (location) (T6), Ella finds that the majority of neighboring circles (over $60\%$ of displayed circles) in the chain-level relationship-view are related to the three relationships previously marked as useful.
These circles form a visual cluster.
Ella checks another circle that is slightly closer to this cluster (T4) than other circles outside the cluster.
\name\ shows detailed connections with both the three different views and the two relationship-views (between person and location, and between person and organization).
Ella finds that it links to a different circle in the person-location relationship-view (T6) and brings more locations and organizations (T5).
Then Ella checks a few more outside circles but none of them are linked to the three relationships marked as useful. 
This causes Ella to think that the three marked relationships may involve coherent or colluded activities.

\textbf{On-demand data retrieval and hypothesis generation.}
To verify this, Ella uses the right-click menu on the three circles to request related documents. 
Based on information in the three marked relationships, \name\ finds and adds relevant documents to Ella's workspace.
For each document, \name\ highlights named entities belonging to the same biclusters (T1). 
This helps direct Ella's attention to useful information in the document.
In reading the text, Ella starts forming a spatialization (Figure \ref{fig-case} \circled{6}) by placing closely related documents near the persons in the graph and manually creating links between the person and some key words in the text (\eg, \texttt{C-4}).
After identifying six key documents and linking them to the graph, Ella thinks that she has enough information.
Using the dragging and automatically moving feature of \name, Ella pulls her created spatialization, including the graph and linked documents, out of the other views.
Referring to this spatialization, Ella hypothesizes that \texttt{Abdul Ramazi} provided the explosive material, \texttt{C-4}, to \texttt{Faysal Goba}, who worked with \texttt{Mukhtar Galab} and \texttt{Yasein Mosed} to plan an attack on \texttt{AMTRAK} train 19.

\section{Initial Expert Feedback}
To better understand the usage of \name, we conducted interviews with two experts:
a data scientist in an IT company (E1) and
a computational scientist of a research institution (E2).   
Both have worked in their fields for over ten years and are experienced in using multiple views for data analysis.
\rv{They were asked to use \name\ to identify potential threats from the Atlantic Storm dataset \cite{hughes2005discovery}.}
It is similar to the usage scenario (Section \ref{scenario}), but has more documents.
After a 30-minute exploration, each gave us feedback on \name.

Overall, they appreciated the capability of seeing computed relationships in stand-alone views.
E1 commented, ``\textit{It [a relationship-view] saves me much effort in the tedious trial-and-error attempts for finding groups of suspicious guys.}"
E2 mentioned, ``\textit{It [a relationship-view] seems working as a summary of related views, so I don't need to check that many nodes in the graph.}"
They benefited from the interactive features offered by \name.
The interactive visual linking was considered useful, with E1 commenting that ``\textit{Following [curved] lines, I can keep browsing related information from one piece to another}", and E2 saying that ``\textit{These [curved] lines help me to track everything.}"
Moreover, they appraised the dragging and automatically moving feature. 
E1 said, ``\textit{It's so convenient that I can move several views at once.}"
E2 mentioned, ``\textit{This well preserves my effort on the layout work.}''

The two experts made three suggestions for improving \name.
First, for visual linking, E2 indicated the need for a recommendation mechanism, by saying ``\textit{While it seems that everything can be linked, it would be better that the system can tell me which paths [several connected elements from multiple views] to check.}"
Second, they both asked for automatic methods to organize multiple views; E1 asked ``\textit{Can it [\name] organize these views?}" and E2 commented, ``\textit{It would be better if the views can be [automatically] ordered.}"
Third, considering the trade-off of using relationship-views, E1 said, ``\textit{Sometimes using them [relationship-views] causes problems, as I had to move a number of views to leave room for them; otherwise they covered others}."

\rv{The feedback suggests the usefulness of \name, especially two of its features: showing computed relationships and interactive visual linking. 
The former saves effort in finding group-level relationships across different views. 
The latter enables users to check relationships in detail.
However, considering that the number of visual links can be large, enabling prioritization of them is helpful to direct user attention to important ones.
Regarding the capability of managing the layout of multiple views, the experts appreciated the function of simultaneously moving multiple views, but their feedback indicates that such a manual way of organizing multiple views is not sufficient.
Although it helps maintain a spatialization of multiple views, manually moving views to create a spatialization is not effective.
This calls for the support of multi-view layout computation.
Moreover, the cost of adding relationship-views was raised.
Thus, whether the benefit of using relationship-views outweighs the cost remains unanswered.
}

\section{Discussion and Conclusion}
We present \name, a visual analytics technique for exploring cross-view data relationships.
\name\ formalizes cross-view data relationships as biclusters and bicluster-chains (for bi-group and multi-group level of relationships) and computes them from data.
\name\ applies a bi-context design that creates stand-alone relationship-views.
Moreover, \name\ allows users to interactively check relationships and visual elements and flexibly organize multiple views.

\rv{\subsection{Comparison to Existing Techniques}

The design of \name\ has three advantages.
First, compared to view-coordination techniques (e.g., brushing and linking), \name\ saves user effort in finding group-level relationships, as they are computed from data and shown in stand-alone relationship-views.
Second, relationship-views separate cross-view data relationships from visual elements in existing views.
This helps reduce the interference with user perception of visual elements in existing views.
Relationship-views also offer an overview that enables users to explore computed relationships directly.
For coordination based techniques, user exploration is limited to visual elements only, as there is no overview of relationships.
Third, compared to simple link techniques (e.g., VisLink \cite{collins2007vislink}), \name\ can show more complex relationships (e.g., group-level relationships).
Moreover, visual elements in relationship-views potentially serve as edge bundles, which reduces the number of edges shown with simple links.
}

We walk through an investigative analysis scenario to demonstrate the usefulness of \name.
While other bicluster-based visual analysis tools (\eg, Bixplorer \cite{sun2014role}, BiSet \cite{sun2016biset} and BiDots \cite{zhao2017bidots}) have also been reported as helpful for such analyses, the key advantages and differences between them and \name\ lie in two aspects.
First, \name\ allows users to browse a variety of views (\eg, map, list, and graph) that enrich visual context for sensemaking activities. 
In Bixplorer, only matrices being offered, and in BiSet and BiDots, users can only rely on lists of information (\eg, searching for Boston from a list of locations in BiSet versus seeing Boston on a map in \name).
\rv{Second, \name\ offers more advanced layout management capabilities than the other three (e.g., dragging and moving a created spatialization that includes several views), which better supports using space to think \cite{andrews2010space}.
BiSet and BiDots support list ordering but do not allow users to shuffle lists, so users can only see linked information in a fixed sequence.}
Bixplorer allows users to move matrices, but is limited to one at a time, so users cannot move a previously created spatialization without breaking it.

\rv{In summary, \name\ can effectively support users in exploring group-level cross-view relationships, as they are computed as biclusters or bicluster-chains from data and displayed in relationship-views.
Such computations do not rely on any application domains, so it is possible to apply \name\ to a variety of analysis scenarios.
With relationship-views, \name\ enables six ways of user exploration (Table \ref{tab-user-explorations}), which cannot be fully supported by existing techniques.
}

\rv{\subsection{Limitations and Future Work}

\name\ computes cross-view data relationships as biclusters. 
Such a calculation may lead to information loss (e.g., not all individual-level relations are included in computed biclusters). 
As relationship-views rely on bicluster computation, individual-level relationships that do not belong to any biclusters cannot be shown.
When no biclusters can be found from data, \name\ behaves the same as the brushing and linking technique, so it performs better for group-level relationships.

Regarding relationship-views in \name, MDS is used to plot computed biclusters.
While it aims to reveal the similarity among biclusters, how effective it is for users to read biclusters and their links to visual elements in the other views is not clear.
This requires further studies.
}

\name\ facilitates exploring and managing multiple views, but with the sacrifice of more visual widgets.
Due to adding relationship-views, it takes more display space and may add more cognitive load to users.
Whether the expected benefits outweigh these risks, or if it is possible to design certain minimal additions, instead of creating additional views, to the existing UI of multiple views, need further exploration.

\rv{Using visual links can cause visual clutter, as discussed in Section \ref{mview-prior-work}, which is another limitation of \name.
As \name\ separates relationships from visual elements, users need visual guidance to check relationships in detail.
To achieve this, \name\ uses curved lines.
However, if there are too many curved lines, especially when users select multiple relationships that involve visual elements from a large number of views, visual clutter occurs.
Thus, similar to using proximity to reveal cross-view data relationships, the curved lines used in \name\ work for small sets of views.
To improve this, we will need to study other visual encodings to reveal visual elements in relationships.
}

\rv{It is possible to extend \name\ to support users in exploring relations between elements of temporal and spatial proximity, which exist in various scenarios. 
The two types of elements are often shown in different views and sometimes they do not exactly match each other.
For such cases, we can choose to compute partial biclusters \cite{eren2013comparative}, which can be easily incorporated into \name.
Moreover, while it is possible to apply \name\ to a variety of data, besides the Sign of the Crescent dataset and the Atlantic Storm dataset,  \name\ currently does not support users in trying their own datasets.
We plan to enhance \name\ with functions of computing partial biclusters and loading different datasets.
}


\acknowledgments{
This research is supported in part by NSF Grants IIS-2002082, SMA-2022443, the Research and Artistry Opportunity Grant from Northern Illinois University, and the NSERC Discovery Grant. 
Maoyuan Sun acknowledges the Argonne Leadership Computing Facility for support during the summer of 2020 and 2021 as part of the Argonne National Laboratory faculty research program.}

\bibliographystyle{abbrv-doi}

\bibliography{paper}
\end{document}